\documentclass[dvips]{article}
\usepackage{icrctc07}

\title{The Monoceros very-high-energy gamma-ray source}
\shorttitle{A point-like $\gamma$-ray source in Monoceros}
\authors{A. Fiasson$^{1}$, J. A. Hinton$^{2}$, Y. Gallant$^{1}$, 
A. Marcowith$^{1}$, O. Reimer$^{3}$, G. Rowell$^{4}$, for the H.E.S.S. Collaboration}
\shortauthors{A. Fiasson et al}
\afiliations{$^1$Laboratoire de Physique Th\'eorique et Astroparticules, IN2P2/CNRS, Universit\'e Montpellier II, CC 70, Place Eug\`ene Bataillon, F-34095 Montpellier Cedex 5, France\\ $^2$School of Physics \& Astronomy, University of Leeds, Leeds LS2 9JT, UK\\$^3$Stanford University, HEPL \& KIPAC, Stanford, CA 94305-4085, USA\\$^4$School of Chemistery \& Physics, University of Adelaide, Adelaide 5005, Australia}
\email{Armand.Fiasson@lpta.in2p3.fr, J.A.Hinton@leeds.ac.uk}

\abstract{The H.E.S.S. telescope array has observed the complex Monoceros Loop SNR/Rosette Nebula region which contains unidentified high energy EGRET sources and potential very-high-energy (VHE) $\gamma$-ray source. We announce the discovery of a new point-like VHE $\gamma$-ray sources, HESS J0632+057. It is located close to the rim of the Monoceros SNR and has no clear counterpart at other wavelengths. Data from the NANTEN telescope have been used to investigate hadronic interactions with nearby molecular clouds. We found no evidence for a clear association. The VHE $\gamma$-ray emission is possibly associated with the lower energy $\gamma$-ray source 3EG J0634+0521, a weak X-ray source 1RXS J063258.3+054857 and the Be-star MWC 148.}

\begin{document}
\maketitle
%Begin the section.

\section{Introduction}
Shell type supernova remnants (SNRs) are believed to be particle accelerator to energy up to a few hundred TeV. Observations of very high energy $\gamma$-ray (VHE; E $\geq$ 100 GeV) from these objects (Aharonian et al. 2006) confirm the presence of particles with energy higher than 10 TeV in these regions. The presence of molecular clouds in the vicinity of SNRs could reveal the nature of such particles as they would interact and produce VHE $\gamma$ rays. The Monoceros SNR (G205.5+0.5), situated at $\sim$1.6 kpc (Graham et al. 1982), apparently interacting with the Rosette Nebula (a young stellar cluster/ molecular cloud complex situated at 1.4 $\pm$ 0.1 kpc(Heinsberger et al. 2000)) is a candidate.\\
In the case of interaction of accelerated particles with interstellar medium producing neutral pions which decays in two $\gamma$ rays, we expect a correlation between $\gamma$-ray emission and matter concentration. We used NANTEN data to trace target material. The NANTEN 4m diameter sub-mm telescope at Las Campanas observatory, Chile, has been conducting a $^{12}$CO (J=1$\rightarrow$0) survey of the galactic plane since 1996, including the Monoceros region (Mizuno \& Fukui 2004).
 
\section{H.E.S.S. observations and results}
\begin{figure}
\begin{center}
\includegraphics [width=0.32\textwidth]{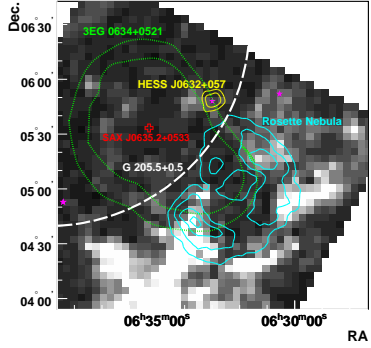}
\end{center}
\caption{$^{12}$CO (J=1$\rightarrow$0) emission from the Monoceros SNR / Rosette Nebula region. The gray-scale corresponds to velocity integrated (0-30 km.s$^{-1}$) emission from the NANTEN Galactic Plane Survey (white areas mean highest flux). The 4$\sigma$ and 6$\sigma$ levels for the statistical significance of a point-like VHE $\gamma$-ray source are shown as yellow contours. Extended cyan contours are radio observations at 8.35 GHz of the Rosette Nebula. The white dashed circle is the Green catalog nominal position and size of the Monoceros SNR. The dotted green contours are 95\% and 99\% confidence level for the position of the EGRET source 3EG J0634+0521. And last, the position of the binary pulsar SAX J0635.2+053 is marked as a red square and the position of Be-stars with pink stars.}
\label{fig1}
\end{figure}

The H.E.S.S. experiment is an array of four Cherenkov telescope installed in Namibia which detects $\gamma$ rays with energy in the 100 GeV to 50 TeV range. A more complete description of the H.E.S.S. experiment is given in Aharonian et al. 2004. The Monoceros loop region has been observed between March 2004 and March 2006 (Aharonian et al. 2007). The dataset includes 13.5~hours of data after quality selection and dead-time correction and was taken at zenith angles ranging between 29$^{\circ}$ and 59$^{\circ}$. It corresponds to a mean energy threshold of 400 GeV with standard cuts used in spectral analysis and 750 GeV with hard cuts used for the source search and position fitting.\\
We made a search for a point-like source on this dataset using a source size of 0.11$^{\circ}$ and a ring of radius 0.5$^{\circ}$ for background estimation. We found an excess corresponding to a statistical significance of 7.1$\sigma$. Fig. \ref{fig1} shows the NANTEN $^{12}$CO map with 4 and 6$\sigma$ levels of statistical confidence contours for the VHE $\gamma$-ray excess (yellow contours). As we made a blind search for a point-like source, the probability we get an excess at this position is increased by the number of positions in the field of view, here $\approx$10$^{5}$. This leads to a post-trials statistical significance of 5.3$\sigma$. The excess is confirmed by an independent analysis based on a fit of camera images to a shower model (\textit{Model Analysis}, see de Naurois 2006), which yields to a significance of 7.3$\sigma$ (5.6$\sigma$ post-trials).\\
\begin{figure}
\begin{center}
\includegraphics [width=0.33\textwidth]{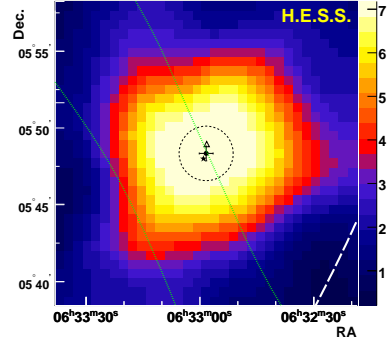}
\end{center}
\caption{Statistical significance map of the H.E.S.S. VHE $\gamma$-ray source. The rms size limit is shown as a dotted circle. Dotted green contours are 95\% and 99\% confidence level for the position of the EGRET source 3EG J0634+0521. The unidentifed X-ray source 1RXS J063258.3+054857 is marked with a triangle and the Be-star MCW 148 with a star.}
\label{fig2}
\end{figure}
The fitted position of this new source HESS J0632+057, is 6$^{h}$32$^{min}$58.3$^{s}$, +5$^{\circ}$48'20" (RA/Dec. J2000) with 28" statistical errors on each axis (fig. \ref{fig2}). We estimated systematics errors at 20" on each axis. The fig. \ref{fig3} represents the distribution of signal in function of the angular distance around the fitted position. The distribution is fully compatible with the point spread function (red curve). We derived an upper limit on the size of the source of 2' at 95\% confidence assuming a Gaussian profile for the source.\\
The reconstructed energy spectrum of the excess is consistent with a power-law of index $\Gamma = 2.53 \pm 0.26 \pm 0.20$ and differential flux at 1 TeV $\Phi_{TeV} = 9.1 \pm 1.7 \pm 3.0 \times 10^{-13} $cm$^{-2}$s$^{-1}$TeV$^{-1}$. The first errors are statistical errors and the second are estimated systematic errors. Fig.\ref{fig4} represents the VHE $\gamma$-ray reconstructed flux together with that for the EGRET sources 3EG0634+0521 and the upper limit derived by the HEGRA telescope array for the EGRET source position (converted to differential flux assuming the spectral shape observed by H.E.S.S.). There is no evidence of flux variability in our dataset but the sparse sampling of data together with the weakness of the source do not permit to constrain strongly intrinsic variability of the source.\\

\begin{figure}
\begin{center}
\includegraphics [width=0.45\textwidth]{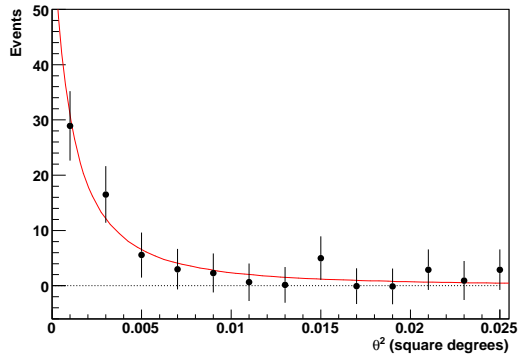}
\end{center}
\caption{Distribution of $\gamma$-ray candidates events as function of squared angular distance from the bes fit position of HESS J0632+057. The red line is the point spread function corresponding to this dataset obtained with Monte-Carlo simulations.}
\label{fig3}
\end{figure}
\begin{figure}
\begin{center}
\includegraphics [width=0.45\textwidth]{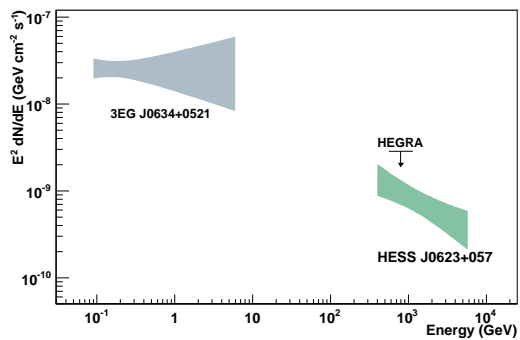}
\end{center}
\caption{Reconstructed VHE $\gamma$-ray spectrum of HESS J0632+057 compared to the EGRET source 3EG J0634+0521. The upper limit obtained using the HEGRA instrument for the EGRET source position is shown.}
\label{fig4}
\end{figure}

\section{Possible associations}
The region where lies HESS J0632+057 is a complex region and although there is no clear counterpart, it may be associated with various objects known at other wavelengths. 
\subsection{3EG J0634+0521}
In the same region lies also an EGRET source, 3EG J0634+0521 (Hartman et al., 1999). Considering that the source is flagged as confused and possibly extended, our measurement, which lies between 95\% and 99\% confidence region, is compatible with its position. Furthermore, the reported third EGRET catalogue flux above 100 MeV is consistent with an extrapolation of the H.E.S.S. spectrum. A global fit of the two spectra gives a photon index of 2.41 $\pm$ 0.06 (fig.\ref{fig4}). 

\subsection{The Monoceros Loop SNR}
The possible association of spectra in the GeV and TeV band is an argument in favor of an hadronic interpretation of the VHE $\gamma$-ray emission. In this case, the Monoceros loop SNR is a good candidate for acceleration of particles. This remnant, which has an age of $\sim$ 3$\times$10$^{4}$ years, is rather old in comparison to known shell type SNRs emitting VHE $\gamma$ rays ($\sim$2000 years). However, cosmic rays acceleration may occur even at later evolutionary phase (late Sedov or Radiative, see Yamazaki et al. 2006). Given the point-like nature of the VHE $\gamma$-ray emission, to explain VHE $\gamma$ rays as a product of accelerated cosmic rays interacting with interstellar medium requires the presence of a dense molecular cloud 
coincident with the emission. An unresolved molecular cloud listed in CO survey at 115 GHz (Oliver et al. 1996) lies rather close to HESS J0632+057. The distance estimate for this cloud (1.6 kpc) is consistent with that for the Monoceros SNR. NANTEN survey shows that the intensity peak of this cloud is significantly shifted to the east of the H.E.S.S. source (fig. \ref{fig1}). There is no evidence of other dense clouds along the line of sight in the NANTEN data.

\subsection{1 RXS J063258.3+054857}
1 RXS J063258.3+054857 is a faint ROSAT source which is potential counterpart of HESS J0632+057, given the uncertainty of the position of the two objects. Given the number of sources in the field of view, the chance probability of coincidence of the two source is 0.1\%. X rays are useful to discriminate between scenario of VHE $\gamma$-ray emission. If $\gamma$-rays are due to inverse Compton scattering from a population of accelerated electrons, X rays are expected to come from synchrotron emission of the same population. In this case, the weakness of this source ($\sim$10$^{-13}$erg cm$^{-2}$ s$^{-1}$) compared to the TeV flux ($\sim$10$^{-12}$erg cm$^{-2}$ s$^{-1}$) required a very low magnetic field ($\ll$3$\mu$G), unless a strong radiation source exists in the neighbourhood of the emission region. Important absorption of the X-ray emission may also explain weakness of the ROSAT source. In the case of a hadronic scenario, production of pions leads to secondary electrons which produce a weaker X-ray source, probably compatible with the measured ROSAT flux.

\subsection{MWC 148}
A massive emission-line Be-star lies within the H.E.S.S. error circle. Given the fact that there are only three stars of this type in the field of view, the chance probability of the association is $\approx$10$^{-4}$. Stars of this spectral type have winds with typical velocities and mass loss rates of 1000 km.s$^{-1}$ and 10$^{-7}$\textit{M}$_{\odot}$. Stellar winds may induce internal or external shocks where particles can be accelerated, but no association of VHE $\gamma$-ray emission with similar stars have been already detected and seems unlikely. Another hypothesis is that this star is a part of a binary system with a compact companion not already detected. Further observations are required to constrain this scenario.

\section{Acknowledgments}
The support of the Namibian authorities and of the University of Namibia in facilitating the construction and operation of H.E.S.S. is gratefully acknowledged, as is the support by the German Ministry for Education and Research, the CNRS-IN2P3 and the Astroparticle Interdisciplinary Programme of the CNRS, the U.K. Particle Physics and Astronomy Research Coucil (PPARC), the IPNP of the Charles University, the South African Department of Science and Technology and National Research Foundation, and by the University of Namibia. We appreciate the excellent work of the technical support staff in Berlin, Durham, Hamburg, Heidelberg, Palaiseau, Paris, Saclay, and in Namibia in the construction and operation of the equipment. The NANTEN project is financially supported from JSPS (Japan Society for the Promotion of Science) Core-to-Core Program, MEXT Grant-in-Aid for Scientific Research on Priority Areas, and SORST-JST (Solution Oriented Research for Science and Technology: Japan science and Technology Agency). We would also like to thank Stan Owocki and James Urquhart for very useful discussions.
\nocite{ref1}
\nocite{ref2}
\nocite{ref3}
\nocite{ref4}
\nocite{ref4b}
\nocite{ref5}
\nocite{ref6}
\nocite{ref7a}
\nocite{ref7}
%This is the reference to .bib file (Whitout .bib!)
\bibliography{icrc0581}
%This in the bibtex style, is ok.
\bibliographystyle{plain}

\end{document}